\documentclass{elsart}

\usepackage{graphicx}

\usepackage{amssymb,amsbsy,amsmath}

\newcommand{\figref}[1]{Fig.~\protect\ref{#1}}
\newcommand{\fmref}[1]{(\protect\ref{#1})}

\newcommand{\dint}{\text{d}}

\journal{Physica A}

\begin{document}
\begin{frontmatter}

  \title{Van Der Waals Revisited}

\author[adr1]{Klaus B\"arwinkel\corauthref{cor1}}
\address[adr1]{Universit\"at Osnabr\"uck, Fachbereich Physik,
D-49069 Osnabr\"uck, Germany}
\corauth[cor1]{Tel: ++49 541 969-2694; fax: -2670; Email: klaus.baerwinkel@uni-osnabrueck.de}

\author[adr2]{J\"urgen Schnack}
\address[adr2]{Universit\"at Bielefeld, Fakult\"at f\"ur Physik,
  Postfach 100131, D-33501 Bielefeld, Germany}
\corauth[cor2]{Tel: ++49 521 106-6193; fax -6455; Email: jschnack@uni-bielefeld.de}

\begin{abstract}
  The van-der-Waals version of the second virial coefficient is
  not far from being exact if the model parameters are
  appropriately chosen. It is shown how the van-der-Waals
  resemblance originates from the interplay of thermal averaging
  and superposition of scattering phase shift contributions. The
  derivation of the two parameters from the quantum virial
  coefficient reveals a fermion-boson symmetry in non-ideal
  quantum gases. Numerical details are worked out for the Helium
  quantum gases.
\end{abstract}

%%%%%%%%%use  the \KEY command at the begin of keyword text%%%%%%%%%

\end{frontmatter}

\section{Occupation number statistics and the van-der-Waals model}
\label{sec-1}

Occupation number statistics for $N=n\Omega$ non-interacting
distinguishable quantum particles (Boltzmann statistics) in a
volume $\Omega$ yields the entropy density functional
%--------------------------------------------------------
\begin{equation}
\label{E-1-1}
s = \frac{k_B}{\Omega} \int \dint^3 p\, \rho(\vec{p})\,
 \nu_{\vec{p}}\, (1-\ln \nu_{\vec{p}})
\end{equation}
%--------------------------------------------------------
where $\nu_{\vec{p}}$ is the average occupation number of a
single-particle energy eigenstate. These eigenstates are
enumerated by corresponding points $\vec{p}$ in momentum space,
the density of which is $\rho (\vec{p})$. $k_B$ denotes
Boltzmann's constant.  For the more general case of
indistinguishable quasi-particles one may consider eq.~(22) of
\cite{BST:PA99}, where the Fermi-Bose functional is
stated. Nevertheless, this simplifies to our eqs.~\fmref{E-1-1} or
\fmref{E-1-5} if $\nu_{\vec{p}}\ll 1$ which is the case that we
discuss in this article.

A hard-core like repulsive interaction is taken into account by
the van-der-Waals ansatz
%--------------------------------------------------------
\begin{equation}
\label{E-1-2}
\rho (\vec{p}) = (\Omega - Nb) / (2\pi \hbar)^3 \ .
\end{equation}
%--------------------------------------------------------
Here the ``single-particle volume'' $b$ is our first
van-der-Waals parameter. We consider elementary cells of volume
%--------------------------------------------------------
\begin{equation}
\label{E-1-3}
v_{e\ell} = \frac{\Omega}{\rho(\vec{p})} = \frac{(2\pi\hbar)^3}{1-nb}
\end{equation}
%--------------------------------------------------------
in six-dimensional phase space ($\mu$-space) and the
one-particle distribution function
%--------------------------------------------------------
\begin{equation}
\label{E-1-4}
f(\vec{p}) = \nu_{\vec{p}} / v_{e\ell}\ .
\end{equation}
%--------------------------------------------------------
The entropy density
%--------------------------------------------------------
\begin{equation}
\label{E-1-5}
s = k_B \int \dint^3 p\, f(\vec{p})\, (1-\ln (f(\vec{p}) v_{e\ell}))
\end{equation}
%--------------------------------------------------------
is then to be maximized as a functional of $f$ subject to the
constraints of fixed particle density
%--------------------------------------------------------
\begin{equation}
\label{E-1-6} 
n = \int \dint^3 p\, f(\vec{p})
\end{equation}
%--------------------------------------------------------
and fixed energy density
%--------------------------------------------------------
\begin{equation}
\label{E-1-7}
u = \int \dint^3 p\, f(\vec{p})\, \epsilon_{\vec{p}}\ .
\end{equation}
%--------------------------------------------------------
According to the second van-der-Waals ansatz, each particle has
its classical kinetic energy and is in the potential field of
interaction with the other particles, i.e.
%--------------------------------------------------------
\begin{equation}
\label{E-1-8}
\epsilon_{\vec{p}} = \frac{p^2}{2m} - a n
\ ,
\end{equation}
%--------------------------------------------------------
where $a$ is the second van-der-Waals parameter. Clearly, the
treatment of correlations is incomplete in this model.

The energy density now becomes
%--------------------------------------------------------
\begin{equation}
\label{E-1-9}
u = \int \dint^3 p\, \frac{p^2}{2m} f(\vec{p}) - an^2\ .
\end{equation}
%--------------------------------------------------------
From the principle of maximum entropy $s$ and using the
temperature definition
%--------------------------------------------------------
\begin{equation}
\label{E-1-10}
T = \left[ \left( \partial s / \partial u \right)_n\right]^{-1}
\end{equation}
%--------------------------------------------------------
one finds $f$ to be the Maxwellian
%--------------------------------------------------------
\begin{equation}
\label{E-1-11}
f(\vec{p}) = n (2\pi m k_B T)^{3/2}\,
 \exp\left\{ - \frac{p^2}{2m k_B T}\right\}
\ .
\end{equation}
%--------------------------------------------------------
This leads to
%--------------------------------------------------------
\begin{equation}
\label{E-1-12}
s = n k_B \left( \frac{5}{2} - \ln \left( \frac{n \lambda^3}{1-nb}\right)\right)
\end{equation}
%--------------------------------------------------------
with the thermal wavelength
%--------------------------------------------------------
\begin{equation}
\label{E-1-13}
\lambda = (2\pi\hbar) / \sqrt{2\pi m k_B T}
\end{equation}
%--------------------------------------------------------
and to
%--------------------------------------------------------
\begin{equation}
\label{E-1-14}
u = \frac{3}{2} n k_B T - an^2\ .
\end{equation}
%--------------------------------------------------------
Then the pressure formula
%--------------------------------------------------------
\begin{equation}
\label{E-1-15}
P_{eq} = - \frac{1}{n^2} \left( \frac{\partial u}{\partial n}\right)_s
\end{equation}
%--------------------------------------------------------
results in the van-der-Waals equation of state:
%--------------------------------------------------------
\begin{equation}
\label{E-1-16}
P_{eq} = \frac{n k_B T}{1-nb} - an^2\ .
\end{equation}
%--------------------------------------------------------
Because of the insufficient treatment of two-particle
correlations, this formula will allow a quantitatively
satisfying fit for real systems only if $n \lambda^3$ is
sufficiently small. Consequently, the van-der-Waals version of
the second virial coefficient $B(T)$ is a good approximation if
the temperature is not too low:
%--------------------------------------------------------
\begin{equation}
\label{E-1-17}
B(T) 
\approx 
B_{\text{vdW}}(T)
=
b - \frac{a}{k_B T}\ .
\end{equation}
%--------------------------------------------------------
The appropriate choice of the parameters $a$ and $b$ is dealt
with in the following sections. In particular, it could very
well be that both parameters depend on the fermionic or bosonic
nature of the interacting particles. It will turn out that this
is not the case.

\section{Heuristics of the van-der-Waals parameters}
\label{sec-2}

The van-der-Waals model can be introduced via corresponding
approximations to the radial distribution function. To this end
consider first the average potential energy of $N = n \Omega$
mutually interacting classical particles:
%--------------------------------------------------------
\begin{equation}
\label{E-1-18}
W_{pot} = \frac{1}{2} <\sum_{i\neq j} V (|\vec{r}_i - \vec{r}_j |>\ .
\end{equation}
%--------------------------------------------------------
With the radial distribution function defined by
%--------------------------------------------------------
\begin{equation}
\label{E-1-19}
g(|\vec{r}^{\ \prime} - \vec{r}^{\ \prime\prime} |> 
= 
n^{-2} \, < \sum_{i\neq j} \delta (\vec{r}^{\ \prime} 
- \vec{r}_i) \delta(\vec{r}^{\ \prime\prime} - \vec{r}_j)>
\end{equation}
%--------------------------------------------------------
one gets
%--------------------------------------------------------
\begin{equation}
\label{E-1-20}
W_{pot} = N \frac{n}{2} \int \dint^3 r\, g(r)\, V(r)
\ .
\end{equation}
%--------------------------------------------------------
Now $g(r)$ has its density expansion
%--------------------------------------------------------
\begin{equation}
\label{E-1-21}
g(r) = g_0 (r) + n g_1(r) + n^2 g_2 (r) + \ldots
\ .
\end{equation}
%--------------------------------------------------------
The energy of a single classical particle is therefore -- apart
from higher-order density contributions -- given by
eq.~\fmref{E-1-8} with 
%--------------------------------------------------------
\begin{equation}
\label{E-1-22}
a = - \frac{1}{2} \int \dint^3 r\, V(r)\, g_0(r)
\ .
\end{equation}
%--------------------------------------------------------
In view of the classical limit
%--------------------------------------------------------
\begin{equation}
\label{E-1-23}
g_{0,cl}(r) = \exp\left\{-\frac{V(r)}{k_B  T}\right\}
\end{equation}
%--------------------------------------------------------
with a Lennard-Jones potential (see below, eq.~\fmref{E-1-35}),
there will be a cut-off radius $r_\ast$ such that the
approximation
%--------------------------------------------------------
\begin{equation}
\label{E-1-24}
g_0 (r) = \left\{ \begin{array}{lrl}
0 & , & \text{ for}\ r < r_\ast\\
1 & , & \text{ for}\ r > r_\ast \end{array}\right.
\end{equation}
%--------------------------------------------------------
with a constant $r_\ast$ is applicable in a considerable range
of temperature. This eventually fixes the parameter $a$ as
%--------------------------------------------------------
\begin{equation}
\label{E-1-25}
a = - \frac{1}{2} \int_{r\ge r_\ast} d^3 r\, V(r)
\ .
\end{equation}
%--------------------------------------------------------
On the other hand, both the parameters $a$ and $b$ may be
introduced by first expressing the second virial coefficient in
terms of $g_0$ \cite{BaG70},
%--------------------------------------------------------
\begin{equation}
\label{E-1-26}
B(T) = \frac{1}{2} \int \dint^3 r\, (1-g_0 (r))
\,
\end{equation}
%--------------------------------------------------------
and employing the closer approximation
%--------------------------------------------------------
\begin{equation}
\label{E-1-27}
g_0 (r) = \left\{ \begin{array}{lrl}
0  & , & \text{ for}\ r < r_\ast\\
1 - \frac{V(r)}{k_B T} & , & \text{ for}\ r > r_\ast
\end{array}\right.
\ .
\end{equation} 
%--------------------------------------------------------
Comparison with the van-der-Waals version of $B(T)$
(eq.~\fmref{E-1-17}) then yields $a$ as given by
eq.~\fmref{E-1-25} and 
%--------------------------------------------------------
\begin{equation}
\label{E-1-28}
b = \frac{2\pi}{3} r^3_\ast
\ .
\end{equation} 
%--------------------------------------------------------
The closer approximation of $g_0$ -- if inserted in
eq.~\fmref{E-1-22} -- would cause a slight dependence of $a$ on
temperature. This must be negligible for the van-der-Waals model
to be acceptable.

In the next section, formulae \fmref{E-1-17}, \fmref{E-1-25},
and \fmref{E-1-28} will be substantiated by numerical analysis
of the exact quantum mechanical virial coefficient, and the
cut-off radius $r_\ast$ will be determined.

\section{The exact virial coefficient and its van-der-Waals limit}
\label{sec-3}

\subsection{Substantiation of the model}

The exact theory \cite{Bau67} for boson or fermion gases with
their two-particle interaction having, possibly, bound state
energies $E_i$ gives the second virial coefficient as a sum of
four terms:
%--------------------------------------------------------
\begin{equation}
\label{E-1-29}
B(T) = 
\mp 2^{-5/2} \lambda^3 - 2^{3/2} \lambda^3 
\sum_i e^{-E_i/k_B T} + \ll G_\pm \gg + \frac{\ll F_\pm \gg}{k_B T}
\end{equation}
%--------------------------------------------------------
with the quantities $\ll G_\pm \gg$ and $\ll F_\pm \gg$ being
explained below (eqs. \fmref{E-1-37}, \fmref{E-1-38}) and with
the upper (lower) sign valid for bosons (fermions). The
double bracket is our notation for the thermal average of
momentum dependent functions, e.g.
%--------------------------------------------------------
\begin{equation}
\label{E-1-30}
\ll \Phi \gg = \int^\infty_0 \dint p\, w(p)\, \Phi(p)\ ,
\end{equation}
%--------------------------------------------------------
with the thermal weight function
%--------------------------------------------------------
\begin{equation}
\label{E-1-31}
w(p) = 4 \pi \, (\pi m k_B T)^{-3/2} \, p^2
\exp\left\{-\frac{p^2}{m k_B T}\right\}
\ .
\end{equation}
%--------------------------------------------------------
Evidently, formula \fmref{E-1-17} is justified if $\ll F_\pm
\gg$ and $\ll G_\pm \gg$ prove to be practically constant,
i.e. $\ll F_\pm \gg = a$ and $\ll G_\pm \gg = b$ in a relevant
range of temperature where the other contributions are
negligible. This is indeed the case as will be shown in the
following. Moreover, $a$ and $b$ thus defined will exhibit a new
kind of fermion-boson symmetry in that they are independent of
the specific quantum statistics.

\begin{figure}[!ht]
\begin{center}
\includegraphics[clip,width=65mm]{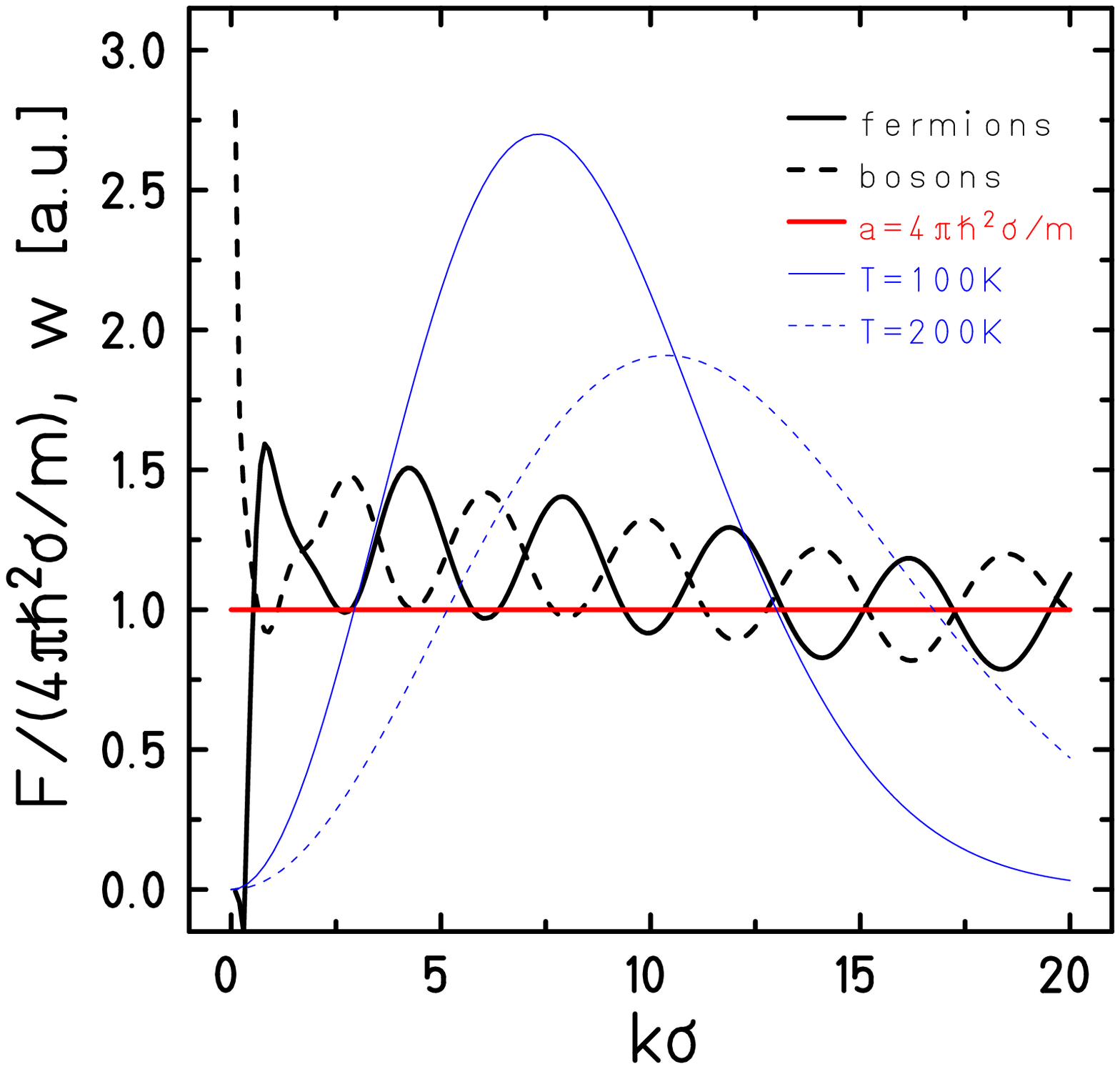}
\quad
\includegraphics[clip,width=65mm]{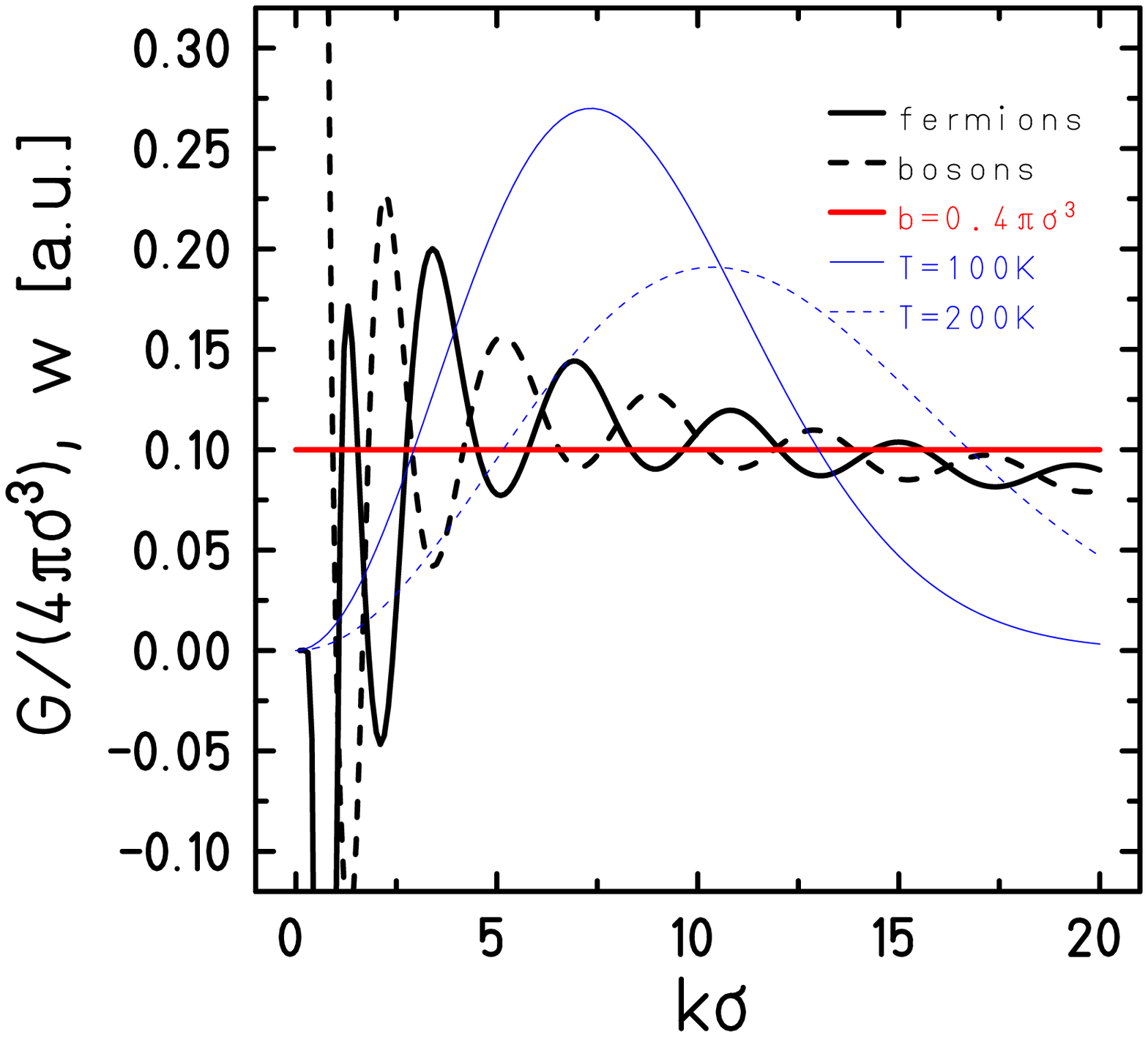}
\vspace*{1mm}
\caption[]{Functions $F_\pm$ and $G_\pm$ for bosons ($^4$He
  atoms) interacting via the Lennard-Jones potential
  \fmref{E-1-35} and fermions with the same mass and same
  interaction. The thermal weight function is given for two
  temperatures. The horizontal line on the l.h.s. marks the
  value of $a$ as given by \fmref{E-1-59} and on the
  r.h.s. the value of $b$ as given by \fmref{E-1-58}.} 
\label{F-1}
\end{center}
\end{figure}

The functions $F_\pm$ and $G_\pm$ may be expressed in terms of
the properly (anti-) symmetrized momentum representation of the
two-particle operator
%--------------------------------------------------------
\begin{equation}
\label{E-1-32}
{\mathcal T} (z) 
= V - V \frac{1}{H-z} V
\ , \
{\mathcal T}^\prime (z) = \frac{d}{dz} {\mathcal T} (z)
\ ,
\end{equation}
%--------------------------------------------------------
with $H = H_{kin} + V$ being the Hamiltonian of relative motion:
%--------------------------------------------------------
\begin{eqnarray}
\label{E-1-33}
F_\pm (p) 
&=& 
- \frac{1}{2} (2\pi\hbar)^3 
\Re\left(< \vec{p}\ | {\mathcal T}_\pm (E_p + i \epsilon>| \vec{p} >\right)
\ , \
E_p = \frac{p^2}{m}
\ ,
\\
\label{E-1-34}
G_\pm (p) &=& \frac{\pi}{2} (2\pi\hbar)^3 \int \dint^3 q\, \delta (E_p - E_q) \cdot\\
          & & \cdot \Im\left( < \vec{p}\ | {\mathcal
          T}_\pm (E_q + i\epsilon) | \vec{q} > 
          < \vec{q}\ | {\mathcal T}_\pm^\prime (E_q + i\epsilon)
          | \vec{p} > \right)
\ .\nonumber
\end{eqnarray}
%--------------------------------------------------------
Our graphics \figref{F-1} for $F_\pm$ and $G_\pm$ rely on the
numerical evaluation for bosons ($^4$He atoms) interacting via a
Lennard-Jones potential lacking bound states \cite{KiK:JCP51}
and fermions (same mass and same interaction as $^4$He):
%--------------------------------------------------------
\begin{equation}
\label{E-1-35}
V(r) = 4 V_0 \left[ \left( \frac{\sigma}{r}\right)^{12} - \left(
  \frac{\sigma}{r}\right)^6 \right]
\ ;\ V_0/k_B = 10.22 \text{K} , \sigma = 2.56 \text{\AA}\ .
\end{equation}
%--------------------------------------------------------
The (anti-) symmetrized ${\mathcal T}$-matrix is -- up to a
multiplicative constant -- nothing else but the scattering
amplitude
%--------------------------------------------------------
\begin{equation}
\label{E-1-36}
f_\pm (p,\theta) = - \pi^2 m \hbar <\vec{p}\ | {\mathcal T}_\pm
(E_p + i \epsilon) | \vec{q} >
\ ,\
| \vec{p} | = | \vec{q} | 
\ ,\
\vec{p} \cdot \vec{q} = p^2 \cos \theta\ .
\end{equation}
%--------------------------------------------------------
An alternative representation for $F_\pm$ and $G_\pm$ can
therefore be given in terms of scattering phase shifts
$\delta_\ell$ which complies with the Beth-Uhlenbeck result for
$B(T)$ \cite{BeU36,BeU37}:
%--------------------------------------------------------
\begin{equation}
\label{E-1-37}
F_\pm = \frac{4\pi\hbar^2}{m} f_\pm (p,0) 
= 
\frac{4\pi\hbar^2}{m} \frac{\hbar}{2p} 
\sum_\ell{}^\pm (2\ell +1) \sin 2 \delta_\ell (p)
\end{equation}
%--------------------------------------------------------
%--------------------------------------------------------
\begin{equation}
\label{E-1-38}
G_\pm (p) = - 4 \pi \hbar \frac{\hbar^2}{p^2} \sum_\ell{}^\pm
(2\ell +1) \sin^2 [\delta_\ell (p)] \frac{\partial \delta_\ell
(p)}{\partial p}\ .
\end{equation}
%--------------------------------------------------------
The summation runs over even $\ell$ for bosons and odd $\ell$ for fermions. 

%===================    figure   =================================
%===================    figure ===============================

A remark on the units and dependencies of $F_\pm$, $G_\pm$ will
be fitting here. Let $p = \hbar k$ and choose $\tilde{r} =
r/\sigma$ as the dimensionless radial coordinate. Then the
dimensionless version of the radial wave equation with
eigenvalue $E_p$ becomes
%--------------------------------------------------------
\begin{equation}
\label{E-1-39}
u_\ell^{\prime\prime} (\tilde{r}) 
+ 
\left\{ \text{Re}\ (\tilde{r}^{-12} -\tilde{r}^{-6}) 
+ \ell(\ell+1)\tilde{r}^{-2}\right\} u(\tilde{r}) 
= (k\sigma)^2 u_\ell (\tilde{r})
\end{equation}
%--------------------------------------------------------
where
%--------------------------------------------------------
\begin{equation}
\label{E-1-40}
\text{Re} = \frac{4 V_0 m \sigma^2}{\hbar^2}
\end{equation}
%--------------------------------------------------------
is the ``Reynolds number''. Given the appropriate behavior of
$u_\ell$ for $\tilde{r} \to 0$, the asymptotic behavior for
$\tilde{r} \to \infty$ exhibits the phase shift $\delta_\ell$:
%--------------------------------------------------------
\begin{equation}
\label{E-1-41}
u_\ell (\tilde{r}) 
\longrightarrow \sin(k \sigma \tilde{r} - \frac{\ell}{2} \pi +
\delta_\ell )
\ .
\end{equation}
%--------------------------------------------------------
Thus the only dependency of $\delta_\ell$ is on Re and
$k\sigma$. Consequently, with some functions $\varphi_\pm = \varphi_\pm
(\text{Re}, k \sigma)$ and $\gamma_\pm = \gamma_\pm (\text{Re},
k\sigma)$
%--------------------------------------------------------
\begin{equation}
\label{E-1-42}
F_\pm = \frac{4\pi\hbar^2 \sigma}{m}\cdot \varphi_\pm (\text{Re},
k\sigma) 
= 16 \pi V_0 \ \sigma^3 \
\frac{\varphi_\pm (\text{Re}, k \sigma)}{\text{Re}}
\ ,
\end{equation}
%--------------------------------------------------------
%--------------------------------------------------------
\begin{equation}
\label{E-1-43}
G_\pm = 4\pi \sigma^3 \ \gamma_\pm(\text{Re}, k\sigma )\ .
\end{equation}
%--------------------------------------------------------
In each of our graphics the boson and the fermion function refer
to the same value of Re. With the potential data of
eq.~\fmref{E-1-35} and with $m$ the mass of $^4$He we obtain 
Re~$\approx 22.1$.

The weight function $w = w(p)$ (eq.~\fmref{E-1-31}) assumes its
maximum at $p_{\text{max}} = \hbar k_{\text{max}}$ with
%--------------------------------------------------------
\begin{equation}
\label{E-1-44}
k_{\text{max}} \sigma 
= \left( \frac{100 \text{Re}}{4\cdot 10,22}\right)^{1/2} 
\cdot \left( \frac{T}{100 K}\right)^{1/2}
\end{equation}
%--------------------------------------------------------
and it is easily seen that above $T=100K$ the weight functions
samples rather large values of $k\sigma$ ($k\sigma \gtrapprox
10$). For high values of $k\sigma$ \figref{F-1} shows that both
$F_+$ and $F_-$ oscillate about a common approximately constant
value. This reflects the nearly hard-core likeness of the
repulsive part of the Lennard-Jones potential. Let us therefore
consider, for comparison, a pure hard-core repulsion with radius
$\sigma$.

\begin{figure}[!ht]
\begin{center}
\includegraphics[clip,width=65mm]{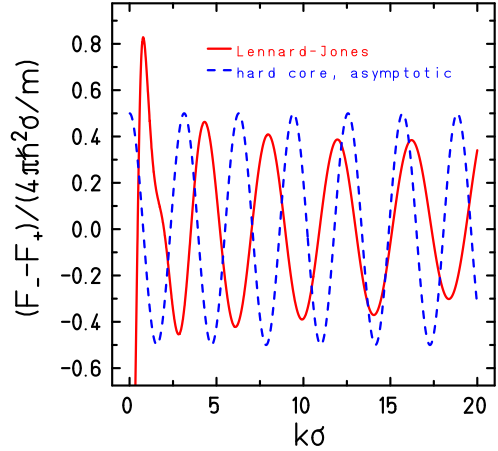}
\vspace*{1mm}
\caption[]{Comparison of $(F_- - F_+)$ for the Lennard-Jones
  system (solid curve) with the asymptotic result for the hard
  core system (dashed curve).}
\label{F-2}
\end{center}
\end{figure}

In this special case
%--------------------------------------------------------
\begin{equation}
\label{E-1-45}
\tan \delta_\ell = j_\ell (k\sigma)/y_\ell (k\sigma)
\end{equation}
%--------------------------------------------------------
holds, with $j_\ell$ and $y_\ell$ denoting the spherical Bessel functions
%--------------------------------------------------------
\begin{equation}
\label{E-1-46} 
j_\ell (z) 
= z^n \left( -\frac{1}{z} \frac{d}{dz}\right)^n 
\frac{\sin z}{z}
\ ,\
y_\ell (z) = z^n \left(\frac{d}{dz}\right)^n \frac{\cos z}{z}
\ .
\end{equation}
%--------------------------------------------------------
Invoking
%--------------------------------------------------------
\begin{equation}
\label{E-1-47}
\sin 2\delta_\ell = 2 \frac{\tan \delta_\ell}{1+\tan^2 \delta_\ell}
\end{equation}
%--------------------------------------------------------
and the asymptotic behavior
%--------------------------------------------------------
\begin{equation}
\label{E-1-48}
\lim_{z\to \infty} 
\left( j^2_\ell (z) + y^2_\ell (z)\right)
= 
1/z^2 
\ ,
\end{equation}
%--------------------------------------------------------
one arrives at
%--------------------------------------------------------
\begin{equation}
\label{E-1-49}
(F_+^{hc} - F_-^{hc})_{asy} 
= \frac{4\pi \hbar^2 \sigma^2 k}{m} 
\sum_{\ell=0}^\infty (-1)^{\ell +1} (2\ell +1) \,
j_\ell(k\sigma)\, y_\ell
(k\sigma)
\ .
\end{equation}
%--------------------------------------------------------
Now -- as a marginal case of formula 10.1.46 in \cite{AbS73} -- 
%--------------------------------------------------------
\begin{equation}
\label{E-1-50}
\sum_{\ell=0}^\infty (-1)^{\ell +1}\, 
(2\ell +1)\, j_\ell(z)\, y_\ell(z) 
= \frac{\cos 2 z}{2\sigma}
\end{equation}
%--------------------------------------------------------
and consequently
%--------------------------------------------------------
\begin{equation}
\label{E-1-51}
(F_-^{hc} - F_+^{hc})_{asy} 
= \frac{2\pi\hbar^2 \sigma}{m} \cos (2k\sigma)\ .
\end{equation}
%--------------------------------------------------------
This is the asymptotic ($k\sigma \to \infty$) result for the
hard-core system. It is compared with $(F_- - F_+)$ for the
Lennard-Jones system in \figref{F-2}. The behavior is very similar
both in terms of amplitude and frequency. A slight decrease of
$\sigma$ with increasing $k$ in formula \fmref{E-1-51} would
still improve the agreement. This reflects the fact that the
Lennard-Jones potential appears the softer the higher the
particles' energy is.

%===================    figure   =================================
%===================    figure ===============================

As for $G_\pm$, an oscillation about a common constant value is
once again seen, compare the r.h.s. of \figref{F-1}. In contrast
to the case of $F_\pm$, however, the oscillatory amplitude of
the difference is clearly decreasing. Not surprisingly, this
feature can again be derived analytically for a pure hard-core
repulsion with radius $\sigma$.

\begin{figure}[!ht]
\begin{center}
\includegraphics[clip,width=65mm]{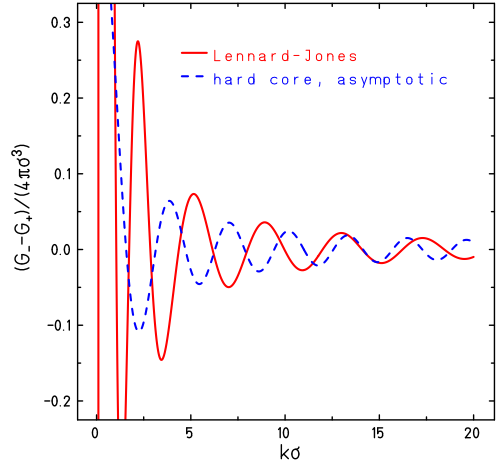}
\vspace*{1mm}
\caption[]{Comparison of $(G_- - G_+)$  for the Lennard-Jones
  system (solid curve) with the asymptotic result for the hard
  core system (dashed curve).}
\label{F-3}
\end{center}
\end{figure}

To this end, $G_\pm$ (eq.~\fmref{E-1-38}) is first rewritten as
%--------------------------------------------------------
\begin{equation}
\label{E-1-52}
G_\pm = \frac{2\pi \sigma^3}{(k\sigma)^2} 
\frac{\partial}{\partial (k\sigma)}  
\sum_\ell{}^\pm (2\ell +1) 
\left( \frac{1}{2} \sin 2 \delta_\ell (k\sigma) - \delta_\ell (k\sigma)\right)
\end{equation}
%--------------------------------------------------------
and then employed for the hard-core system. With the aid of
eqs. \fmref{E-1-45}, \fmref{E-1-47}, and \fmref{E-1-48} the
asymptotic $(k\sigma \to \infty)$ tail of $(G_+^{hc} -
G_-^{hc})$ is found to be
%--------------------------------------------------------
\begin{equation}
\label{E-1-53}
(G_+^{hc} - G_-^{hc})_{asy} 
= 
2\pi \sigma^3 \sum^\infty_{\ell =
0}\, (-1)^\ell\, (2\ell +1)\, j_\ell(k\sigma)\, y_\ell^\prime
(k\sigma)\ .
\end{equation}
%--------------------------------------------------------
After replacing (for $k\sigma \to \infty$) $y^\prime_\ell$ by
$j_\ell$, one can apply formula 10.1.51 in \cite{AbS73},
%--------------------------------------------------------
\begin{equation}
\label{E-1-54}
\sum^\infty_{\ell =0}\, 
(-1)^\ell (2\ell +1) j_\ell^2 (z) = \frac{\sin 2
  z}{2z}
\ ,
\end{equation}
%--------------------------------------------------------
and hence
%--------------------------------------------------------
\begin{equation}
\label{E-1-55}
(G_+^{hc} - G_-^{hc})_{asy} 
= 2\pi \sigma^3 \frac{\sin 2k\sigma}{2k\sigma}
\ .
\end{equation}
%--------------------------------------------------------
The difference $(G_+ - G_-)$ for the Lennard-Jones system is
compared with this result in \figref{F-3}. Again, the behavior
is very similar both in terms of amplitude and frequency, also
the phase difference increases only slightly.

%===================    figure   =================================
%===================    figure ===============================

\subsection{The cut-off radius}

Instead of eq.~\fmref{E-1-27} a continuous ansatz for $g_0 (r)$
may be used: 
%--------------------------------------------------------
\begin{equation}
\label{E-1-56}
g_0 (r) = \left\{ \begin{array}{lrl}
0 & , & \text{ for} \ r\le_\ast\\
1 - \exp \left(-\alpha \frac{r-r_\ast}{r_\ast}\right) 
+ \frac{V(r) - V(r_\ast) 
\exp(-\alpha \frac{r-r_\ast}{r_\ast})}{k_B T} 
& , & \text{ for}\ r \ge r_\ast\end{array}\right.
\end{equation}
%--------------------------------------------------------
with the potential $V(r)$ according to eq.~\fmref{E-1-35}. 
Equation~\fmref{E-1-27} is reproduced for $\alpha \rightarrow
\infty$ . The
values of $r_\ast/\sigma$ and $\alpha$ follow from the
van-der-Waals condition
%--------------------------------------------------------
\begin{equation}
\label{E-1-57}
b - \frac{a}{k_B  T} = \frac{1}{2} \int \dint^3 r (1-g_0(r))
\end{equation}
%--------------------------------------------------------
where
%--------------------------------------------------------
\begin{equation}
\label{E-1-58}
b = 0.4 \pi \sigma^3
\end{equation}
%--------------------------------------------------------
and
%--------------------------------------------------------
\begin{equation}
\label{E-1-59}
a = \frac{4\pi \hbar^2 \sigma}{m} = \frac{16\pi}{\text{Re}} \cdot V_0 \sigma^3
\end{equation}
%--------------------------------------------------------
is estimated in view of \figref{F-1} (horizontal lines). Then
%--------------------------------------------------------
\begin{equation}
\label{E-1-60}
r_\ast = x^{-1/3}\sigma
\end{equation}
%--------------------------------------------------------
where $x$ is the solution of 
%--------------------------------------------------------
\begin{equation}
\label{E-1-61}
x = \frac{3}{\text{Re}} + 0.3 x^2 + \frac{2}{3} x^3 - 0.3 x^4
\end{equation}
%--------------------------------------------------------
with $0.2\ x > 1/3$. The parameter $\alpha$ follows from
%--------------------------------------------------------
\begin{equation}
\label{E-1-62}
0.2 x - 1/3 
= \frac{1}{\alpha} + \frac{2}{\alpha^2} + \frac{2}{\alpha^3}
\ .
\end{equation}
%--------------------------------------------------------
Then $a$ has the following representation in terms of the
Reynolds number Re and the Lennard-Jones parameters $V_0$ and
$\sigma$:
%--------------------------------------------------------
\begin{equation}
\label{E-1-63}
a = \frac{4 \pi V_0 \sigma^3}{3}
\left(
x - \frac{1}{3} x^3
\right)
\ ,
\end{equation}
%--------------------------------------------------------
where $x=x(\text{Re})$ is that one among the solutions of
eq.~\fmref{E-1-61}, which allows a positive $\alpha$ in
eq.~\fmref{E-1-62}.  For Re$=22.1$ we obtain $x=1.88$ and
$\alpha=25.6$. Therefore, in our ansatz \fmref{E-1-56} $g_0(r)$
is close to its limit for $\alpha \rightarrow\infty$, which is
given in eq.~\fmref{E-1-27}.

\section*{Summary}

The summary of our elaboration is that for a large region of
temperature the functions $\ll F_\pm \gg = a$ and $\ll G_\pm \gg
= b$ may be nearly considered as constants, i.e.~not depending
on temperature, see \figref{F-1}. Moreover, the respective kind
of statistics (Bose-Einstein or Fermi-Dirac) does not
matter. This constitutes a fermion-boson symmetry in non-ideal
quantum gases. For ideal quantum gases such symmetries are known
for about ten years \cite{Lee:PRE97A,ScS:PA99}.

The slight residual dependence of $a$ and $b$ on temperature
(cf. \figref{F-1}) reflects the fact that these are parameters
not of an exact but of a model theory.

\section*{Acknowledgment}

A long-standing collaboration with our colleague and friend
Heinz-J{\"u}rgen Schmidt is gratefully acknowledged. This
article is dedicated to him on the occasion of his 60th
birthday.

%%%%%%%%%%%%%%%%%%%%%%%%%%%%%%%%%%%%%%%%%%%%%%%%%%%%%%%%%%%%%%%%%%%%%% 

%\bibliographystyle{/home/schnack/tex/bibtex/bst/elsart-num.bst}
%\bibliography{/home/schnack/tex/bibtex/js-own,/home/schnack/tex/bibtex/js-mag,/home/schnack/tex/bibtex/js-mis}

\begin{thebibliography}{1}
\expandafter\ifx\csname url\endcsname\relax
  \def\url#1{\texttt{#1}}\fi
\expandafter\ifx\csname urlprefix\endcsname\relax\def\urlprefix{URL }\fi

\bibitem{BST:PA99}
K.~B{\"a}rwinkel, J.~Schnack, U.~Thelker, Quasi-particle picture for monatomic
  gases, Physica A 262 (1999) 496.

\bibitem{BaG70}
K.~B\"arwinkel, S.~Gro{\ss}mann, Pair distribution function of moderately dense
  quantum fluids, Z. Phys. 230 (1970) 141.

\bibitem{Bau67}
B.~Baumgartl, Second and third virial coefficient of a quantum gas from
  2-particle scattering amplitude, Z. Phys. 198 (1967) 148.

\bibitem{KiK:JCP51}
J.~E. Kilpatrick, M.~F. Kilpatrick, Discrete energy levels associated with the
  {L}ennard-{J}ones potential, J. Chem. Phys. 19 (1951) 930.

\bibitem{BeU36}
G.~Uhlenbeck, E.~Beth, The quantum theory of the non-ideal gas. {I}. deviations
  from the classical theory, Physica 3 (1936) 729.

\bibitem{BeU37}
G.~Uhlenbeck, E.~Beth, The quantum theory of the non-ideal gas. {II}. behaviour
  at low temperatures, Physica 4 (1937) 915.

\bibitem{AbS73}
M.~Abramovitz, I.~Stegun (Eds.), Handbook of Mathematical Functions, Dover, New
  York, 1973.

\bibitem{Lee:PRE97A}
M.~H. Lee, Equivalence of ideal gases in two dimensions and {L}anden's
  relations, Phys. Rev. E 55 (1997) 1518.

\bibitem{ScS:PA99}
H.-J. Schmidt, J.~Schnack, Thermodynamic fermion-boson symmetry in harmonic
  oscillator potentials, Physica A 265 (1999) 584.

\end{thebibliography}

\end{document}